\documentclass[twocolumn,trackchanges]{aastex63}

\received{xx xx, 2021}
\revised{xx xx, 2021}
\accepted{xx xx, 2021}
\submitjournal{ApJ}

\shorttitle{Radio SED of J0100+2802}
\shortauthors{Liu et al.}

\begin{document}

\title{Exploring the radio spectral energy distribution of the ultraluminous radio-quiet quasar SDSS J0100+2802 at redshift 6.3}

\correspondingauthor{Ran Wang}
\email{rwangkiaa@pku.edu.cn}

\author[0000-0001-9321-6000]{Yuanqi Liu}
\affiliation{Department of Astronomy, School of Physics, Peking University, Beijing 100871, P. R. China}
\affiliation{Kavli Institute for Astronomy and Astrophysics, Peking University, Beijing 100871, P. R. China}

\author{Ran Wang*}
\affiliation{Kavli Institute for Astronomy and Astrophysics, Peking University, Beijing 100871, P. R. China}

\author{Emmanuel Momjian} 
\affiliation{National Radio Astronomy Observatory, P.O. Box O, Socorro, NM 87801, USA}

\author{Jeff Wagg}
\affiliation{SKA Observatory, Lower Withington Macclesfield, Cheshire SK11 9FT, UK}

\author{Xiaolong Yang}
\affiliation{Shanghai Astronomical Observatory, Key Laboratory of Radio Astronomy, CAS, 80 Nandan Road, Shanghai 200030, China}

\author{Tao An}
\affiliation{Shanghai Astronomical Observatory, Key Laboratory of Radio Astronomy, CAS, 80 Nandan Road, Shanghai 200030, China}

\author{Yali Shao}
\affiliation{Max-Planck-Institut für Radioastronomie, Auf dem Hügel 69, 53121 Bonn, Germany}

\author{Chris L. Carilli} 
\affiliation{National Radio Astronomy Observatory, P.O. Box O, Socorro, NM 87801, USA}
\affiliation{Cavendish Laboratory, University of Cambridge, 19 J. J. Thomson Avenue, Cambridge CB3 0HE, UK}

\author{Xuebing Wu}
\affiliation{Department of Astronomy, School of Physics, Peking University, Beijing 100871, P. R. China}
\affiliation{Kavli Institute for Astronomy and Astrophysics, Peking University, Beijing 100871, P. R. China}

\author{Xiaohui Fan}
\affiliation{Steward Observatory, University of Arizona, 933 North Cherry Avenue, Tucson, AZ 85721, USA}

\author[0000-0003-4793-7880]{Fabian Walter}
\affil{Max Planck Institute for Astronomy, K\"onigstuhl 17, 69117 Heidelberg, Germany}
\affil{National Radio Astronomy Observatory, P.O. Box O, Socorro, NM 87801, USA}

\author{Linhua Jiang}
\affiliation{Department of Astronomy, School of Physics, Peking University, Beijing 100871, P. R. China}
\affiliation{Kavli Institute for Astronomy and Astrophysics, Peking University, Beijing 100871, P. R. China}

\author{Qiong Li}
\affiliation{Department of Astronomy, School of Physics, Peking University, Beijing 100871, P. R. China}
\affiliation{Kavli Institute for Astronomy and Astrophysics, Peking University, Beijing 100871, P. R. China}
\author{Jianan Li}
\affiliation{Department of Astronomy, School of Physics, Peking University, Beijing 100871, P. R. China}
\affiliation{Kavli Institute for Astronomy and Astrophysics, Peking University, Beijing 100871, P. R. China}
\author{Qinyue Fei}
\affiliation{Department of Astronomy, School of Physics, Peking University, Beijing 100871, P. R. China}
\affiliation{Kavli Institute for Astronomy and Astrophysics, Peking University, Beijing 100871, P. R. China}
\author{Fuxiang Xu}
\affiliation{Department of Astronomy, School of Physics, Peking University, Beijing 100871, P. R. China}
\affiliation{Kavli Institute for Astronomy and Astrophysics, Peking University, Beijing 100871, P. R. China}

\begin{abstract}

We report deep Karl G. Jansky Very Large Array (VLA) observations of the optically ultraluminous and radio-quiet quasar SDSS J010013.02 + 280225.8 (hereafter J0100+2802) at redshift $z=$6.3. We detected the radio continuum emission at 1.5~GHz, 6~GHz, and 10~GHz. This leads to a radio power-law spectral index of $\alpha = -0.52\pm0.18$ ($S \propto \nu^{\alpha}$). The radio source is unresolved in all VLA bands with an upper limit to the size of $0.2^{\prime \prime}$ (i.e., $\sim$ 1.1 kpc) at 10~GHz. We find variability in the flux density (increase by $\sim 33\%$) and the spectral index (steepened) between observations in 2016 and 2017. We also find that the VLA 1.5~GHz flux density observed in the same year is 1.5 times that detected with the Very Long Baseline Array (VLBA) in 2016 at the same frequency. This difference suggests that half of the radio emission from J0100+2802 comes from a compact core within 40 pc, and the rest comes from the surrounding few kpc area which is diffuse and resolved out in the VLBA observations. The diffuse emission is four times brighter than that would be expected if driven by star formation. We conclude that the central active galactic nucleus is the dominant power engine of the radio emission in J0100+2802.

\end{abstract}


\section{Introduction} \label{sec1}

Quasars at $z \ge 6$ can serve as special laboratories to explore the formation and evolution of the first supermassive black holes (SMBHs), which formed at the end of cosmic reionization \citep{Fan2006, Mazzucchelli2017}. There are more than 200 quasars discovered at $z\ge 6$ (e.g. \citealt{Willott2005, Willott2010, Jiang2009, Venemans2015b, Matsuoka2016}). SDSS J010013.02+280225.8 (hereafter J0100+2802) at $z=$6.327, discovered 
by \citet{Wu2015}, has a bolometric luminosity of $4.29 \times 10^{14} L_{\odot}$ and hosts a SMBH with an estimated mass of $1.3 \times 10^{10} M_{\odot}$. This quasar is the highest redshift object among the optically ultraluminous quasars with $M_{1450}\leq-29$ known to date (\citealt{Onken2020, Schindler2021}). 
With detected X-ray emission from $XMM-Newton$ (0.2$-$10 keV) and $Chandra$ (2$-$10 keV) observations, J0100+2802 follows the $\alpha_{OX}-L_{\rm 2500 \AA}$ relation of quasars with comparable ultraviolet luminosity at low to median ($1<z<5$) redshifts \citep{Ai2017}. The dust continuum, $\rm [CII]$, and CO emission lines have been mapped with multi-band ALMA observations at sub-kpc scale resolution \citep{Wang2019}. The far-infrared (FIR) dust continuum indicates a nuclear star formation rate (SFR) of $1900 \ M_{\odot} \ \rm yr^{-1}$, while high-\textit{J} CO line emission is detected as evidence of Active Galactic Nucleus (AGN) contribution to heating the gas and dust in its host galaxy. Therefore, J0100+2802 provides a unique chance to study the early growth and nuclear activities of the most massive SMBHs in the epoch of cosmic reionization. 

J0100+2802 is categorized as a radio-quiet quasar, with the definition of radio loudness $R<10$, where the radio loudness is defined as the ratio of the rest-frame 5~GHz (radio) to the $4400\AA$ (optical) flux densities $R = S_{\rm 5 GHz}/S_{\rm 4400 \AA}$ \citep{Kellermann1989}. But it is not silent with an observed flux density of $104.5 \pm 3.1 \rm \mu Jy$ at 3~GHz \citep{Wang2016}. Although radio-quiet quasars account for $\sim 90\%$ of the optically selected quasars \citep{Ivezic2002, Kellermann2016, Banados2015, Liu2021}, their radio emission properties are not well studied at the highest redshift due to limited sensitivity. Therefore, little is known about their radio spectral index and emission mechanisms. The current observational studies of high redshift quasars are concentrated on radio-loud ones, especially the blazars (e.g.,\citealt{Zhang2017,Zhang2020,An2020}). 

Only a compact core with a flux density lower than expected is detected with the Very Long Baseline Array (VLBA) at milliarcsecond (mas) resolution at 1.5~GHz \citep{Wang2017}. Measuring the radio spectral energy distribution (SED) with the VLA at arcsec resolution would help to determine if the radio spectrum is indeed flattened at low frequency, or if there is undetected radio emission on a large (kpc) scale, which is crucial to explore the origins of the radio emission. 

In this work, we carry out VLA observations of J0100+2802 at 1.5, 6, and 10~GHz and derive a power-law spectral index. This paper is organized as follows: the new radio observations and data reduction are described in Section \ref{sec2}. The observational results and analysis are shown in Section \ref{sec3}. We discuss the missing emission and possible origins in Section \ref{sec4}. The main conclusions are briefly summarized in Section \ref{sec5}. 
 
 We adopt a $\Lambda$CDM with $\rm H_0 = 70 \ km \ s^{-1} Mpc^{-1}$, $\Omega_{\rm m} = 0.3$ and $\Omega_{\Lambda} = 0.7$. With this cosmological model, 1$\arcsec$ at the redshift of J0100+2802 ($z=6.3$) corresponds to 5.5 kpc.  Throughout the article, we model the rest-frame radio continuum emission with a power law of $S_{\nu} \propto \nu^{\alpha}$, where $S_{\nu}$ is the flux density at frequency $\nu$ and $\alpha$ is the spectral index. All the frequencies mentioned are in the observed frame, unless noted otherwise.
 
 \begin{figure*}
\gridline{\fig{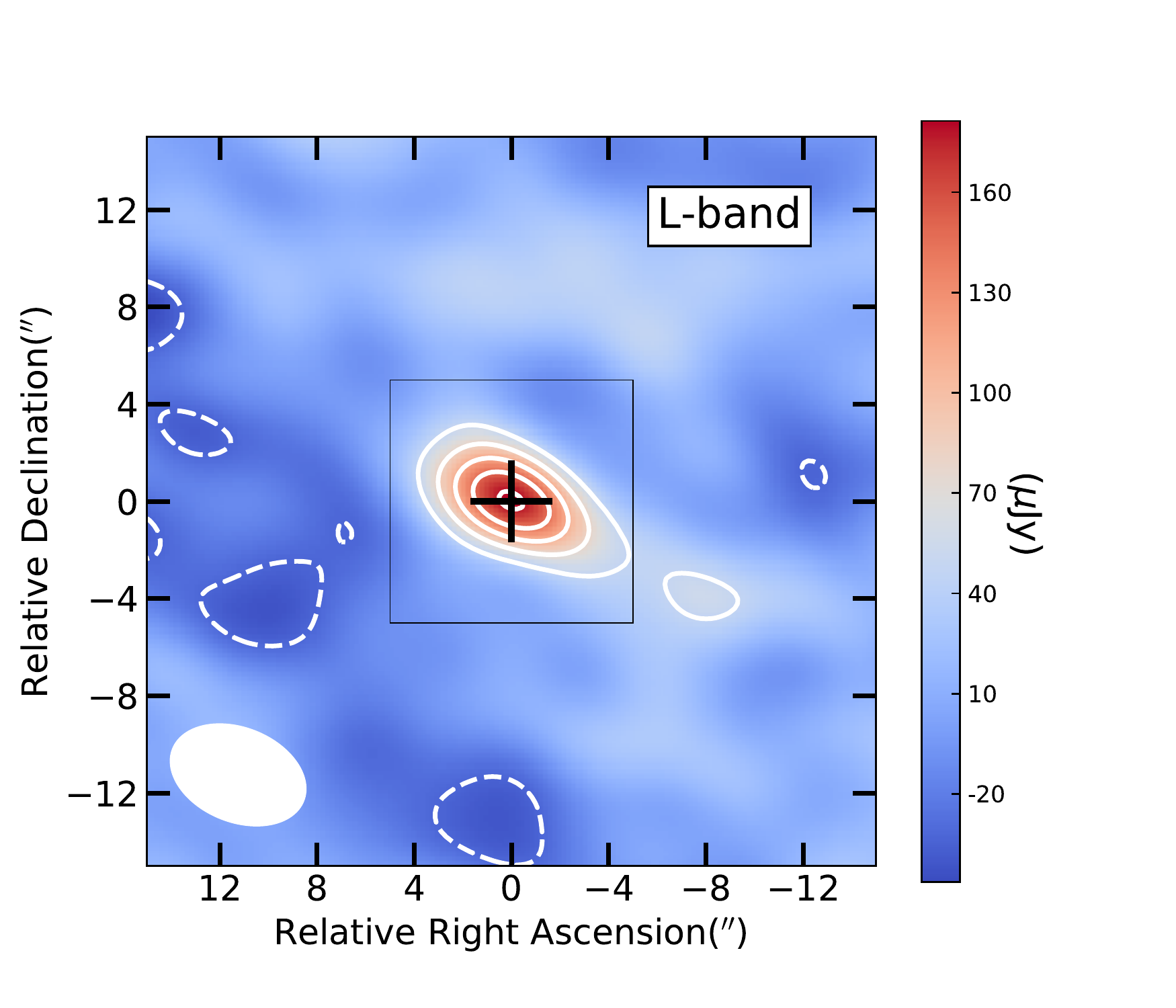}{0.35\textwidth}{(a)}
          \fig{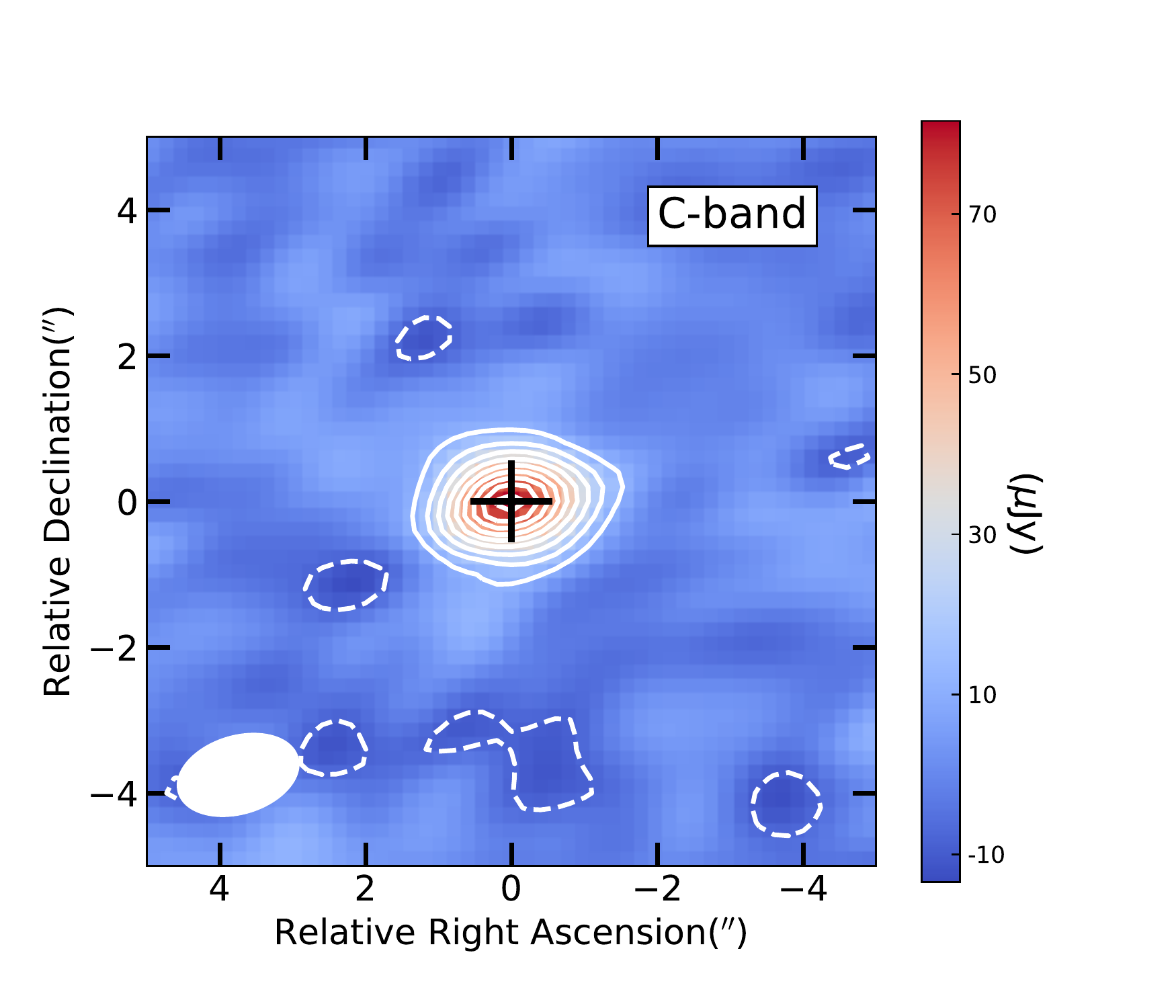}{0.35\textwidth}{(b)}
          \fig{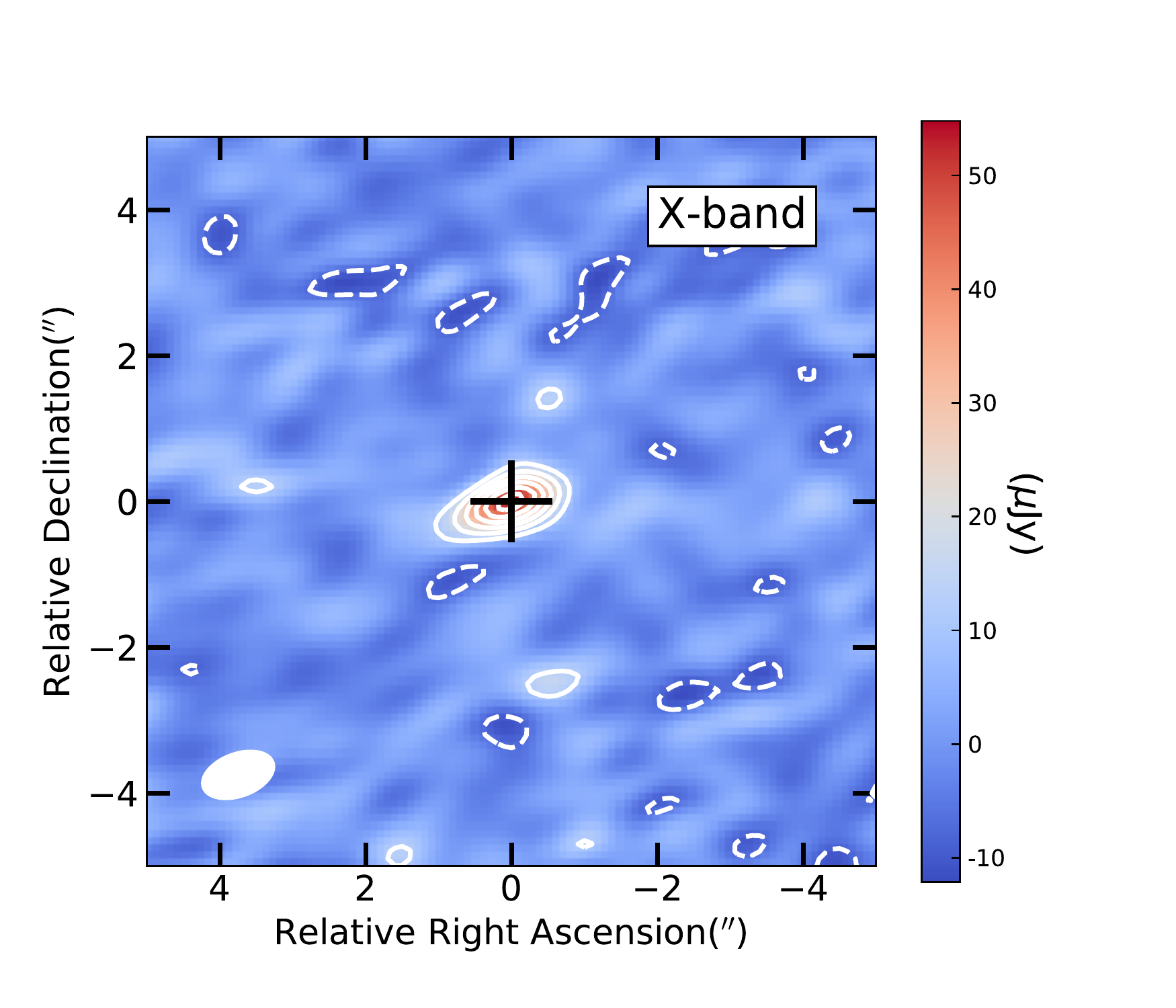}{0.35\textwidth}{(c)}
          }
\caption{J0100+2802 images at (a) 1.5~GHz, (b) 6~GHz and (c) 10~GHz VLA observations. The contour levels are at ($-$2, 3, 5, 7, ..., 19) times the corresponding $1\sigma$ rms noise level of each image (see Table \ref{tab1}). Black crosses denote the optical position (RA = 01h00m13.027s, Dec = +28\arcdeg02\arcmin25.80\arcsec). Synthesized beams are shown as the ellipses at the bottom left. The box in panel (a) shows the image size of panels (b) and (c).
\label{1}}
\end{figure*}

\begin{table*}
\centering
\caption{New and archive radio deep observations of J0100+2802.}
\begin{tabular}{cccc}
\hline
\hline
  & L-band  & C-band & X-band  \\[6pt]
\hline
Observing frequency range (GHz)  & 1.0 $-$ 2.0 & 4.0 $-$ 8.0 & 8.0 $-$ 12.0 \\ 
Observing central frequency (GHz)  & 1.5& 6.0 & 10.0  \\ 
Date of observations (in 2017) &$1^{st}$ Nov &$16^{th}$ Sep &$07^{th}$ Sep \\
Total on-source time (min)        & 100     & 40    &  40   \\ 
Phase calibrator   &J0119+3210   &J0057+3021    &J0057+3021      \\  
Flux/Bandpass calibrator   &  3C48& 3C48   &3C48      \\  
FWHM beam size* ($\arcsec$) & 5.8$\times$3.8 & 1.7$\times$1.1 &1.0$\times$0.6 \\
Beam P.A. ($^{\circ}$) &68.3 & $-$73.4 & $-$70.0 \\
Image $1\sigma$ rms sensitivity ($\mu \rm Jy \  beam^{-1}$) & 16 & 4 &  4 \\
Peak flux density ($\mu \rm Jy \  beam^{-1}$) & 181 & 82&  55 \\
Flux density of tapered image ($\mu \rm Jy$) &  &$86\pm 7$ & $69 \pm 7$ \\[8pt]
\hline  
VLBA observations in 2016: & & & \\
Date of observations & Feb-Mar &  & \\
Total flux density ($\mu \rm Jy$) & $88\pm 19$ & & \\[4pt]
VLA observations in 2016: & & & \\
Observing frequency range (GHz)  & 1.0 $-$ 2.0 & 4.5 $-$ 6.5 & \\ 
Observing central frequency (GHz)  & 1.5 & 5.5 &  \\ Date of observations & May-Aug & Jan & \\
FWHM beam size* ($\arcsec$) & 4.6$\times$3.6 & 4.9$\times$3.7 &   \\
Flux density ($\mu \rm Jy$) & $ 136\pm  10$ & $ 104\pm  5$ & \\
\hline  
\hline  
\end{tabular}
\renewcommand\arraystretch{1.6}
\begin{flushleft}
*note: using robust weighting for observations both in 2016 and in 2017.
\end{flushleft}
\label{tab1}
\end{table*}

\section{observations and data reduction} \label{sec2}
We observed the radio continuum emission from J0100+2802 using the VLA in L, C, and X bands, spending a total of 4 hours in September and November 2017 in B-configuration. The observations employ phase-referencing as our target is faint. The L band observations use the 8-bit samplers, while C and X band observations use the 3-bit samplers.

Data reduction was performed with the Common Astronomy Software Applications package (CASA) version 5.6.3 \citep{McMullin2007} using the standard VLA calibration pipeline. Data affected by radio frequency interference (RFI) were removed, and the imaging was performed using Briggs weighting with a robust parameter of 0.5. The flux density calibration accuracy is about $5\%$ in all three bands. Self-calibration cannot improve the images because the target is not strong enough.  The observation parameters and results are summarized in Table \ref{tab1}. 

J0100+2802 was also observed at 1.5~GHz (L-band) and 5~GHz (C-band) in 2016 (program 16A-242), and the results are published in \citet{Sbarrato2021}. The L-band observation in 2016 was carried out in B configuration, and the C-band observations was in C configuration. We re-calibrated and re-imaged the data from this program with the same procedure as described above for the 2017 VLA observations. We also collect the observations from other VLA frequencies and VLBA \citep{Wang2016,Wang2017}, and low frequency data from LOFAR \citep{Gloudemans2021}.  These results are summarized in Table \ref{tab1}. We compare these measurements in Section \ref{sec3.2}.

\section{results and analysis} \label{sec3}

\subsection{Radio emission and spectral index}
The quasar J0100+2802 is detected in all three bands with signal to noise ratio (SNR) larger than 10. The total intensity maps are shown in Figure \ref{1}. The source is unresolved in all three bands, hence we simply use the peak surface brightness in each band as the representative value of the continuum flux density.We adopt the uncertainty in the flux density measurement from the rms noise in the image because calibration errors are negligible in comparison. The continuum flux densities at 1.5~GHz, 6~GHz, and 10~GHz are $181\pm16 \ \mu \rm Jy$, $82 \pm4 \ \mu \rm Jy$, and $55 \pm4 \ \mu \rm Jy$, respectively (Table \ref{tab1}). The positions of the radio center and optical center from the SDSS catalog are consistent within the uncertainties of the astrometry accuracy of SDSS. 

To compare the flux densities at the same scale, we tapered the visibility data of the 6~GHz and 10~GHz observations to a similar resolution of the 1.5~GHz image. There are no significant differences ($> 2\sigma$) in flux density between full resolution images and tapered images at both 6~GHz and 10~GHz. We will adopt the results of the tapered images at 6~GHz and 10~GHz with the same resolution as the 1.5~GHz image for analysis that follows. These values are listed in Table \ref{tab1}.

\begin{figure}
\epsscale{1.2}
\plotone{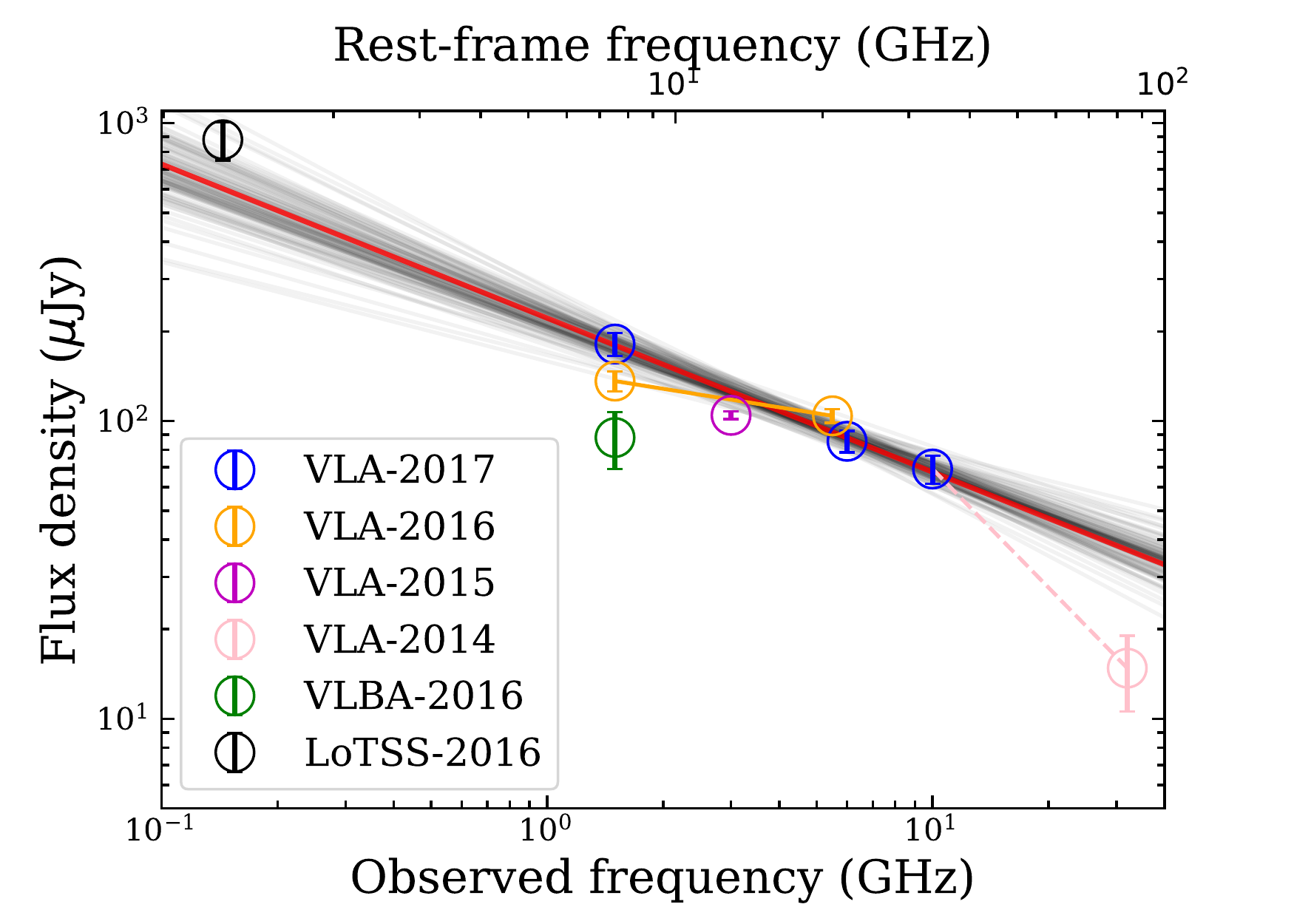}
\caption{Radio continuum of J0100+2802 at observing frequencies of 1.5 to 32~GHz at different epochs. The new VLA observations at 1.5~GHz, 6~GHz, and 10~GHz in 2017 are shown as blue circles, which (with their error bars) are used in fitting the power-law model by the MCMC algorithm. Their errors contain only map rms and not calibration errors. Grey lines represent 100 sets of models that are randomly selected from parameter space, which visualize the uncertainties. The red line denotes the best fitting result with a spectral index of $\alpha = -0.52\pm0.18$. For comparison, the orange circles refer to VLA 1.5~GHz and 5~GHz flux densities observed in 2016 (program 16A-242), which indicate a two-point spectra of $\alpha_{2016} = -0.21\pm0.10$. The green circle shows the flux density from the VLBA 1.5~GHz observations in 2016 \citep{Wang2017}, the magenta circle represents the VLA 3~GHz in 2015, and the pink circle represents the 32~GHz flux density in 2014 \citep{Wang2016}. The pink dashed line shows a possible steeper spectra fitted from X-band and Ka-band observations. 
\label{2}}
\end{figure}

We fit a power law to the radio continuum using the VLA measurements at 1.5, 6, and 10~GHz. 
We derive the spectral index by applying the MCMC algorithm with the Python package $\texttt{emcee}$ \citep{Foreman2012}. The 100 models in parameter space (grey) and best-fitting result of the spectral index (red) are shown in Figure \ref{2}. The continuum flux densities are shown as blue circles. The radio emission measured with the VLA can be well fitted by a single power-law with a spectral index of $\alpha = -0.52\pm0.18$. There is no sign of flattening or turnover in the rest-frame frequency range of 7 to 84~GHz. 

A mean spectral index of $\alpha \sim -0.5$ is reported from multiple investigations using large radio-quiet quasar samples at low to medium redshift ($0<z<3$, e.g., \citealt{Barvainis1996, Kukula1998, Zajacek2019, Gim2019}).
The spectral index of $\alpha = -0.52\pm0.18$ found in J0100+2802 is within the range of spectral indices in low$-z$ radio-quiet quasars. This suggest common radio emission origins and mechanisms, e.g.,  synchrotron emission powered by AGN \citep{Bridle1984, Laing2013} and/or nuclear star forming activity \citep{Condon1992, Condon2013}. 
We will further discuss the radio emission mechanism in Section \ref{sec4.2}. 

With the derived spectral index, we estimate the radio loudness ($R$) of J0100+2802 to be $R=0.6$, confirming the result in \citet{Wang2016} that the source is radio quiet (i.e., $R\leq 10$, \citealt{Kellermann1989, Sikora2007}).

\subsection{Comparison with radio continuum from other epochs}\label{sec3.2}
\begin{figure}
\epsscale{1.2}
\plotone{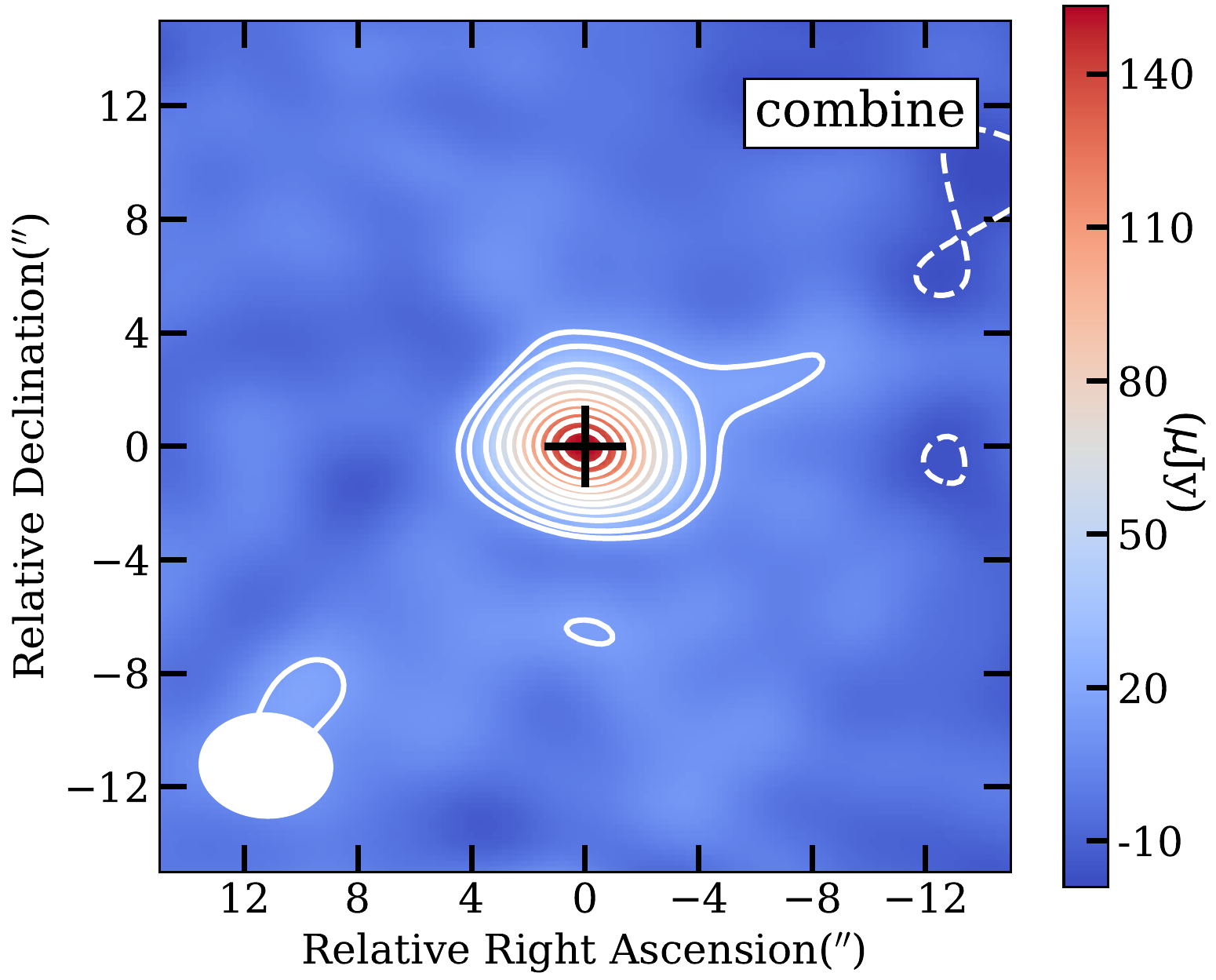}
\caption{J0100+2802 images combining the observations of 2016 and 2017 at 1.5~GHz. The contour levels are at ($-$2, 2, 3, 5, 7, ..., 19) times the corresponding $1\sigma$ rms noise level of 7 $\mu$Jy/beam. The synthesis beam is shown as left bottom ellepse with a FWHM size of $4.6^{\prime\prime} \times 3.6^{\prime\prime}$.
\label{3}}
\end{figure}

The VLA observations in 2016 show an unresolved source with peak flux density of $136.2\pm 10.4 \ \rm \mu Jy$ with a beam size of $4.6^{\prime \prime} \times 3.6^{\prime \prime}$ at 1.5~GHz and a flux density of $104.1\pm 5.4  \ \rm \mu Jy$ with a beam size of $4.9^{\prime \prime} \times 3.7^{\prime \prime}$ at 5.5~GHz. 
The fit to the data in 2016 yields a power-law spectral index of $\alpha_{2016} = -0.21\pm0.10$.
We note that the L-band measurement we obtained is slightly lower than the value of $154\pm 12 \ \rm \mu Jy$ published in \citet{Sbarrato2021}, but is still marginally consistent considering the uncertainties. This is likely to be due to the difference in calibration and imaging methods. Although a tentative jet is reported in \citet{Sbarrato2021}, there is no sign of extended structures with SNR$>$3 in our 1.5~GHz and 6~GHz images. Combining the data from 2016 and 2017 to achieve better sensitivity, no signal of jet emission is detected at $\leq 3\sigma$ in the image as shown in Figure \ref{3}.

The deep VLA S-band (3~GHz) and Ka-band (32~GHz) observations of this source were reported in \citet{Wang2016}, with continuum flux densities of $S_{\rm 3 \ GHz} = 104.5 \pm 3.1 \ \mu \rm Jy$ and $S_{\rm 32 \ GHz} = 14.8\pm  4.3\ \mu \rm Jy $, respectively. These two detections are shown as magenta circles in Figure \ref{2}. The flux density measured at 3~GHz is consistent with both the power-law spectra fitted from observations in 2017 and those in 2016.  
As shown in the figure, the 32~GHz flux density does not fit this single power-law well. Thermal emission may also contribute to the flux density at such high frequency in the source rest frame, so the current detection should be the upper limit of radio activity. This may suggest a steeper spectral index of $\alpha = -1.32\pm 0.31$ at the higher frequencies, i.e., at rest-frame $>$ 100~GHz to 234~GHz. 
However, the observed-frame 32~GHz continuum flux density has large uncertainties due to the low SNR ($\sim 3$). 
The quasar J0100+2802 was also detected in the Low Frequency Array (LOFAR) Two Metre Sky survey (LoTSS-DR2, \citealt{Shimwell2022}) which is published very recently \citep{Gloudemans2021}. The LoTSS data was taken in October 2016 at 144~MHz with a point source flux density of $S_{\rm 144 \ MHz} = 0.88 \pm 0.13 \ \mu \rm Jy$. The LOFAR measurement, combined with the VLA L-band data in 2017 reveals a steep spectral index of $\alpha_{low1} = 0.67 \pm 0.10$  or $\alpha_{low2}=0.80\pm0.10$ using the 2016 data, which suggest that the radio spectrum is getting steeper toward the lower frequency.
 
The VLBA observations in 2016 of J0100+2802 shows a compact core with no extended structures. The intrinsic brightness temperature is estimated to be $T_B = (1.6\pm 1.2)\times 10^7 \ \rm K$, which is higher than the typical value of normal galaxies ($\leqslant 10^5 \ \rm K$, \citealt{Condon1992}). Thus, the VLBA emission likely is due to non-thermal synchrotron emission driven by AGN activity rather than a star formation origin, such as from compact HII regions, radio SNe, or compact SN remnants (See \citealt{Maini2016} for more details). 

\section{discussion} \label{sec4}
\subsection{Variability and the missing emission} \label{sec4.1}
A comparison of the VLA measurements in 2016 and 2017 suggestion possible variability in both radio flux densities and spectral index. The flux density at 1.5~GHz is increased by $45.1 \pm 19.2 \ \mu \rm Jy$ ($33\%$ of the total flux in 2016) in a time period of 15 months in the observed frame. This period corresponds to 2 months in the rest frame. Meanwhile, the radio emission is marginally decreased at 5GHz by $9.7 \pm 9.5 \ \mu \rm Jy$, deriving the 5~GHz flux density in 2017 with the fitted spectral index. Thus, the spectral index fitted between these two bands has steepened from $\alpha_{2016} = -0.21\pm0.10$ to $\alpha_{2017} = -0.54\pm0.13$. 

The VLBA observations were carried out in 2016 February and March, which is 4 months in the observed frame on average before the VLA 1.5~GHz observations in 2016 May to August. We find a 1.5~GHz flux density difference of $48.2 \pm 21.6 \mu \rm Jy$ ($35\%$ of the total flux detected by VLA in 2016) between the VLA and VLBA observations. 
This may also be understood as an increase of radio emission as was suggested by the two-epoch VLA observations at 1.5~GHz. However, we note that the VLBA measurement is lower than the flux density extrapolated from the 3~GHz data observed in 2015 as well \citep{Wang2016}. Thus, in addition to radio variability, radio emission on different scales also need to be seriously considered to explain the difference in flux density between the VLA and VLBA flux densities.

\subsection{Emission on different scales} \label{sec4.2}

The minimum resolvable size in our VLA observations is $1.1 \arcsec$, which is estimated from $\theta = b_{\phi}\sqrt{\rm{\frac{4ln2}{\pi}ln(\frac{SNR}{SNR-1})}} $, with beam size $b_{\phi}=\sqrt{b_{max}\times b_{min}} =4.1 \arcsec$ \citep{Kovalev2005}.  This angular size corresponds to 6.1 kpc at the redshift of J0100+2802. Adopting the same formula, we obtain the minimum resolvable size of VLA 6~GHz and 10~GHz observations in 2017, which are 1.6 kpc and 1.1 kpc, respectively. As the source flux densities in the tapered images only marginally increased within the uncertainties ($<1\sigma$ in 6~GHz, $2\sigma$ in 10~GHz results), we consider the radio emission detected in these three bands to be from the same region. This implies that the size of the radio source detected in our VLA observation is on a scale of $\lesssim 1.1$ kpc.

The VLBA mas-scale resolution imaging of the central region yields a source size of $(40\pm20)\times(18\pm10)$ pc and a flux density of $88.0 \pm 19.0 \ \mu \rm Jy$ \citep{Wang2017}. 


The radio flux that is missed by the VLBA observation is likely to arise from the central kpc region and that is unresolved by the VLA. Such a component could be too extended to be detected by the VLBA (e.g., a few hundreds pc to 1 kpc) and/or consist of several radio knots with surface brightness below the VLBA detection limit.

The missing radio flux density problem was widely reported with mas-resolution imaging of samples of radio-quiet quasars, while the origins of such diffuse emission are rarely discussed. For the radio-quiet Palomar-Green (PG) quasars with VLA observations, a certain percentage of them would be detected by VLBI observations with a lower flux densities (e.g., 8/12, \citealt{Blundell1998}, 9/16, Wang in prep). For moderate redshift ($1<z<2$) radio-quiet quasars, around $50\% \sim 75\%$ flux densities are recovered by VLBI observations \citep{Ruiz2016,Maini2016}. Not only radio-quiet quasars, also radio-loud quasars and other types of sources can show missing flux on VLBI scales (e.g., the radio-loudest quasar at redshift 6, \citealt{Momjian2018}, four hot dust-obsecured galaxies \citealt{Frey2016}). 

In radio-quiet quasars, origins of the radio emission can be both star formation and AGN activity, such as weak jets, winds, and free-free emission from the corona \citep{White2017, Panessa2019}. For example, high-resolution VLBI observations reveal a prominent jet and multiple-phase outflows in Mrk 231, which is the closest ($z=0.042$) radio-quiet AGN \citep{Neff1988,Ulvestad1999a, Ulvestad1999b,Wang2021}. 
Therefore, the missing radio emission of J0100+2802 can be powered by star formation in the surrounding area, or the central engine, or a combination of both. We discuss the possible origins in the following Sections.

\subsubsection{Emission from star formation} \label{secSF}
Firstly, we assume that all the radio emission which is undetected at 1.5~GHz by the VLBA originates from star formation. The corresponding SFR is estimated to be $7900 \ M_{\odot} \ \rm yr^{-1}$, according to the relation in \citet{Murphy2011}:
\begin{equation} \label{eq1}
\left(\frac{\rm{SFR_{1.4 \ GHz}}}{ \rm M_{\odot} \ yr^{-1}}\right) = 6.35 \times 10^{-29} \left(\frac{L_{\rm 1.4 \ GHz}}{\rm erg \ s^{-1} Hz^{-1}}\right) ,
\end{equation}
which is based on the FIR-radio correlation of IR luminous star-forming galaxies \citep{Yun2001}. Here we use a typical radio spectral index of $\alpha = -0.7$ for star forming galaxies to derive the luminosity $L_{\rm 1.4 GHz}$ at rest-frame 1.4~GHz \citep{Condon2013}.
If we assume a nuclear SFR of $1900 M_{\odot} \ \rm yr^{-1}$ \citep{Wang2019} in the central kpc region, based on $equation$ \ref{eq1}, the rest-frame 1.4~GHz radio luminosity associated with the star-forming activity is $L_{\rm 1.4 GHz} = 1.1\times10^7 L_{\odot}$. Note that the dust-continuum based SFR should be considered as an upper limit as AGN activity could contribute to the dust heating. Adopting the same radio spectral index of $\alpha = -0.7$, this corresponds to a flux density of $11.6 \ \rm \mu Jy$ at 1.4~GHz in the observing frame. This accounts for only around $1/4$ of the VLA radio flux density not detected by the VLBA. Therefore, nuclear star formation is unlikely to be the dominant source of the radio emission on kpc scales.

\subsubsection{Emission from AGN activity} \label{secAGN}
AGN outflows on $>$100 pc to kpc scale, including uncollimated winds or low-surface brightness collimated jets, could be a possible source of the radio emission in radio-quiet quasars \citep{Panessa2019}.

Radio emission originating from winds has been reported in \citet{Hwang2018} with a sample of radio-quiet quasars at $z=2-3$. Winds driven by AGNs are normally identified by narrow-line kinematics, for instance, the blueshifted $[\rm O \ III]$ emission line \citep{Christopoulou1997, Zakamska2014, Zakamska2016}. 
The evidence of AGN-powered outflow in J0100+2802 is ambiguous. The quasar's broad MgII emission is blue-shifted by $1020\pm 250 \ \rm km \ s^{-1}$ compared to the [CII] and CO redshift of the quasar host galaxy, suggesting outflowing gas in the broad line region. However, there is no evidence of kpc-scale outflows based on the ALMA observations of [CII] and CO lines in the central kpc region of the quasar host galaxy in 2016 (\citealt{Wang2019}). The narrow-line kinematics is impossible to measure for the quasar J0100+2802. The wind can be inferred from the high optical luminosity, which is sufficient to produce synchrotron emission from shock accelerated relativistic electrons ($\nu L_{\nu} \sim 10^{-5}L_{\rm AGN}$, $L_{\rm AGN}$ is AGN bolometric luminosity \citep{Ishibashi2011, Nims2015}.)

Weak, low-surface brightness jets driven by AGN can be another possible origin of the missing radio emission in radio-quiet quasars. Strong jets are highly collimated, while less powerful jets may have a lower efficiency of collimating the flow \citep{Falcke1995,Panessa2019}. 

In the recent study of the hyperluminous, dust-obscured, and radio-quiet quasar W2246-0526 at redshift $z=4.6$ \citep{Fan2020}, the authors also discuss the AGN-driven origins of the missing radio emission. VLBI observations
resolve a compact core of flux density $75\pm9 \ \rm \mu Jy$, which only accounts for $10\%$ of the total flux density observed with the VLA at the same frequency. Similar to J0100+2802, the flux density difference represents low-brightness diffuse emission at $\geq 32$ pc (VLBI observed component size) to kpc scale. They consider such VLBI-undetected radio emission as being associated with pc-scale winds and/or a low-power jet. The VLBA-undetected radio emission of J0100+2802 may have similar origins. 

\subsection{Overview of the missing emission} \label{secsum}
We consider that both time variability and emission from different scales may contribute to the discrepant flux density between the VLA and VLBA measurements at 1.5~GHz. While the compact radio core emission detected by both VLA and VLBA is likely to be dominated by the central luminous AGN \citep{Wang2017,Sbarrato2021} the radio emission at larger scale maybe from a more diffuse component powered by AGN wind and/ or low-surface brightness knots of weak jets. Nuclear star forming activity could also contribute to the diffuse radio emission on kpc scale.

J0100+2802 is an example to show the missing emission in radio-quiet quasars at $z>6$. It provides a new window to explore the origin and mechanisms of radio emission at different scales. The related AGN-driven wind shocks and the low-power jet are considered to provide a feedback mode in radio-quiet quasars \citep{King2015,Ishibashi2021}.
High sensitivity VLA and VLBI observations of a larger sample are needed, to better constrain the percentage of missing emission and understand the mechanisms. The radio variability in J0100+2802 should also be confirmed with additional VLA and VLBA observations.

\section{conclusion} \label{sec5}

From studies of quasars in the Universe, we now know that the radio-quiet sources should be the most abundant population. However, at the highest redshifts, this faint population could not be studied in detail due to the limits of observational sensitivity. J0100+2802 is a particular optically luminous and radio-quiet quasar at $z\ge 6$, that has been detected with both high-resolution VLBA observations and a relatively complete radio SED covering from the rest-frame frequency range 7 to 234~GHz. 

In this paper, we report the radio spectral index to be $\alpha = -0.52\pm0.18$ fitted from the multi-wavelength radio continuum flux densities observed in 2017. We find a slight variability from 2016 to 2017, which possibly indicates AGN activity.
We compare the VLA 1.5~GHz result with VLBA observations at the same frequency \citep{Wang2017} to constrain the radio emission at different spatial scales. The compact core accounts for more than half of the emission, which is triggered by AGN activity. The missing emission is resolved on tens of pc to kpc scales. This diffuse emission may originate from star formation and AGN-driven outflow. Overall, we conclude that the radio emission is mainly due to AGN rather than star formation. J0100+2802 is currently {an excellent} example of a radio-quiet quasar at such high redshift. However, the appearance of low-power jet and AGN wind on kpc scale are difficult to distinguish. Deep imaging in the radio at intermediate ($\sim 0.1^{\prime \prime}$) resolution filling the gap between VLA and VLBA resolutions will be critical to address this question.


To better understand the characteristics of radio SEDs of the most distant quasars, we require deeper and high-resolution observations of a complete sample of radio-quiet quasars. From research on spatial distribution of radio emission as well as other radio features (e.g. morphology, spectral index), the respective contribution of AGN and star formation activity will be better constrained. 
The missing radio emission may help better understand the AGN feedback and the evolution of SMBHs.

\vspace{5mm}

\facilities{VLA, CASA}

\acknowledgments
This work is supported by National Key Program for Science and Technology Research and Development (grant 2016YFA0400703). We acknowledge the supports from the National Science Foundation of China (NSFC) grants No.11721303, 11991052, 11373008, 11533001. We thank Professor Gregory J. Herczeg, PhD student Lulu Zhang and Yang Li in PKU for writing comments in this article. The National Radio Astronomy Observatory (NRAO) is a facility of the National Science Foundation operated under cooperative agreement by Associated Universities, Inc. This paper makes use of the VLA data from program 17B-071, 16A-242.




\bibliography{sample63}{}
\bibliographystyle{aasjournal}

\end{document}